\documentclass[%
reprint,
superscriptaddress,
letter,
 amsmath,amssymb,
 aps,
longbibliography
]{revtex4-1}

\usepackage{graphicx}% Include figure files
\usepackage{dcolumn}% Align table columns on decimal point
\usepackage{bm}% bold math
\begin{document}

\preprint{APS/123-QED}

\title{Optically Controlled Stochastic Jumps of Individual Gold Nanorod Rotary Motors}% Force line breaks with \\

\author{Lei Shao}
\altaffiliation[Also at ]{Department of Physics, The Chinese University of Hong Kong, Shatin, Hong Kong SAR, China}
\email{shaolei@cuhk.edu.hk}
\affiliation{ Department of Physics, Chalmers University of Technology, S-412 96 G\"oteborg, Sweden}
\author{Daniel Andr\'en}
\affiliation{ Department of Physics, Chalmers University of Technology, S-412 96 G\"oteborg, Sweden}
\author{Steven Jones}
\affiliation{ Department of Physics, Chalmers University of Technology, S-412 96 G\"oteborg, Sweden}
\author{Peter Johansson}
\altaffiliation[Also at ]{Department of Physics, Chalmers University of Technology, S-412 96 G\"oteborg, Sweden}
\affiliation{School of Science and Technology, \"Orebro University, S-701 82 \"Orebro, Sweden}
\author{Mikael K\"all}
\email{mikael.kall@chalmers.se}
\affiliation{ Department of Physics, Chalmers University of Technology, S-412 96 G\"oteborg, Sweden}

\date{\today}% It is always \today, today,

\begin{abstract}
Brownian microparticles diffusing in optical potential energy landscapes constitute a generic testbed for nonequilibrium statistical thermodynamics and has been used to emulate a wide variety of physical systems, ranging from Josephson junctions to Carnot engines. Here we demonstrate that it is possible to scale down this approach to nanometric length-scales by constructing a tilted washboard potential for the rotation of plasmonic gold nanorods. The potential depth and tilt can be precisely adjusted by modulating the light polarization. This allows for a gradual transition from continuous rotation to discrete stochastic jumps, which are found to follow Kramers dynamics in excellent agreement with stochastic simulations. The results widen the possibilities for fundamental experiments in statistical physics and provide new insights in how to construct light-driven nanomachines and multifunctional sensing elements.
%\begin{description}
%\end{description}
\end{abstract}

\pacs{Valid PACS appear here}% PACS, the Physics and Astronomy
%Use showkeys class option if keyword
\maketitle

%\tableofcontents

\section{\label{sec:level1}Introduction}

Laser tweezing is a powerful noninvasive tool to control and measure the movement of Brownian colloidal particles for applications in biology, physics, and chemistry \cite{Ashkin:86,ashkin1987optical,Ashkin1517,moffitt2008recent}. By tuning the light field distribution in the illuminated region, it is possible to generate a wide variety of optical potential energy landscapes. A thermally-driven Brownian particle trapped in such a landscape represents an ideal model system for studying a wide range of fundamental phenomena. Examples to date include studies of chemical reactions \cite{kramers1940brownian,RevModPhys.62.251}, protein folding \cite{neupane2016direct}, thermodynamic relations \cite{wang2002experimental}, information flow \cite{berut2012experimental}, entropy production \cite{tietz2006measurement} and Kramers-type dynamics \cite{rondin2017direct}. By modulating optical potentials, it has even been possible to realize optical Brownian ratchets \cite{wu2016near} and single-particle microscopic Carnot engines \cite{martinez2016brownian}. Interested readers can refer to \cite{C6SM00923A} to find more examples of using optically trapped colloidal particles to build stochastic heat engines.

The tilted periodic ``washboard'' potential is a particularly important potential distribution realizable using optical tweezers technology because it can be used as an archetypal nonequilibrium model of statistical physics in a variety of systems, such as the damped pendulum, ring-laser gyroscope, Josephson junctions, superionic conduction, phase-locked loops, and charge-density-wave condensation \cite{rice2013advances, Risken1996, fulde1975problem, stratonovich1967topics}. Despite the comprehensive theoretical discussion of Brownian dynamics in a tilted washboard-type potential \cite{Risken1996,coffey2004langevin}, it is still of great interest to achieve convenient and precise shape control of such optical potentials. Moreover, it is of generic interest to be able to scale down the physical dimension of the probe particle from the micron range to the nanoscale in order to approach length scales relevant to molecular interactions and dynamics. In addition, nanoscale Brownian particles have much shorter characteristic diffusion time than their micrometer counterparts and thus allow for faster and more efficient sampling of thermodynamic transitions between different states.

In this article, we employed elliptically polarized laser tweezers to create a tilted washboard rotational potential for trapping of colloidal gold nanorods \cite{chen2013gold}. The optical anisotropy and enhanced light-matter interaction of such particles, caused by plasmon resonances \cite{lehmuskero2015laser,shao2018light}, result in extremely efficient optical confinement and rotation performance \cite{shao2018light,friese1998optical}. By adjusting the depth and tilt of the potential by control of polarization ellipticity, we successfully managed to switch the rotational movement of a nanorod from ultrafast continuous spinning to discrete stochastic rotational jumps. Using both experiments and simulations, we further investigated the jump dynamics at critical trapping polarizations and found that it quantitatively agrees with that predicted from Kramers theory \cite{kramers1940brownian,RevModPhys.62.251}. The full control of Brownian rotation of plasmonic nanorods demonstrated here provides an additional freedom of nanomotor movement manipulation and holds great potential for future investigations of fundamental questions in non-equilibrium thermodynamics. Our experimental configuration might also be useful for studies of molecular motors, optical Brownian ratchets, and optical torque wrenches for high-sensitivity biological experiments \cite{pedaci2011excitable}.

\section{Results}

\subsection{\label{sec:level2}Construction of a rotational tilted washboard potential}

An elliptically polarized plane wave with electric field $\bm{E}=E_0[\cos(\omega t)\widehat{x}+\cos(\omega t+\Delta \phi)\widehat{y}]$ can be decomposed into one linearly polarized and one circularly polarized component, $E_\text{L}$ and $E_\text{C}$, respectively (Fig.~\ref{figure1}a, a detailed analysis is provided in the Supplementary Material \footnote{See Supplemental Material at http://link.aps.org/supplemental/ for more details and discussions related to this letter.}). Once a gold nanorod is optically trapped (Fig.~\ref{figure1}b), the linear component provides a restoring torque $M_\text{L}$ that tends to align it along the corresponding polarization direction while the circular component induces a torque $M_\text{C}$ that tends to spin the particle around the direction of incidence to overcome the potential barrier formed by $M_\text{L}$ (Fig.~\ref{figure1}c) \cite{ruijgrok2011brownian}. Both torques are determined by angular momentum transfer due to light absorption as well as scattering \cite{shao2018light,shao2015gold}. The response of the gold nanorod is approximately that of a dipole with induced moment $\bm{p}=\bm{\alpha}\cdot\bm{E}$. For an incident wavelength close to the longitudinal plasmon resonance of the particle, the polarizability tensor $\bm{\alpha}$ is dominated by the long axis component \cite{chen2013gold}, which we denote $\alpha$. The optical potential experienced by the nanorod then has a “tilted washboard” shape (Fig.~\ref{figure1}d) according to:
\begin{equation}
U(\varphi)=-\int_{0}^{\varphi}(-M_\text{L}+M_\text{C})d\varphi=-A\varphi+B\sin^2\varphi
\label{eq:one}
\end{equation}
where $A=1/2\cdot\text{Re}(\alpha) E_\text{C}^2$, $B=1/2\cdot\text{Re}(\alpha) E_\text{L} ^2$ and $\varphi$ is the angle between the nanorod long axis and the linear polarization component $E_\text{L}$ (Fig.~\ref{figure1}d inset). It is easily shown that each well in $U(\varphi)$ is surrounded by highly asymmetric barriers when $\Delta\phi$, the phase difference between the $\widehat{x}$ and $\widehat{y}$ field-components, is below 45$^\circ$. For example, we find that the plasmonic nanorod has to overcome a barrier height $\Delta U=0.8 k_\text{B}T$ to rotate towards the preferred direction while the barrier in the opposite direction is more than an order of magnitude higher for parameter settings mimicking our experimental conditions. The probability of rotational jumps in the ``wrong'' direction is thus very low. The barrier height $\Delta U$ can be varied by changing the degree of polarization ellipticity \textit{via} a change in $\Delta\phi$.
\begin{figure}
\includegraphics{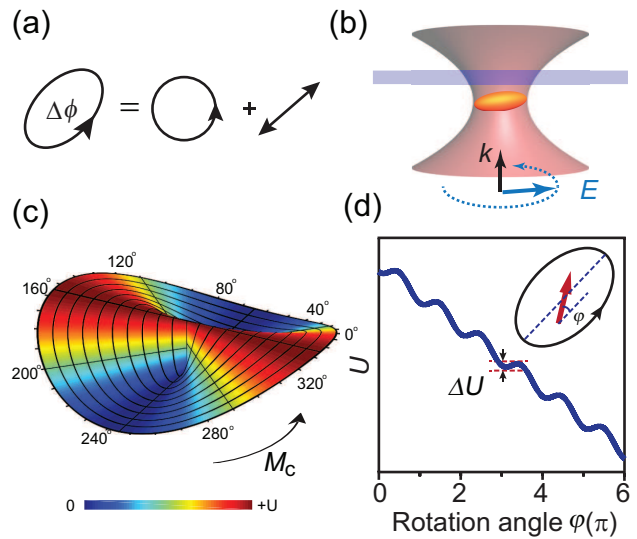}% Here is how to import EPS art
\caption{\label{figure1} Tilted washboard potential from elliptically polarized light. (a) An elliptically polarized plane wave can be decomposed into one linear and one circular polarization component. (b) Schematic of a gold nanorod in an optical trap. The nanorod is oriented with its long axis perpendicular to the direction of incidence. (c) Nanorod potential energy distribution induced by the linear polarization component of an elliptically polarized light field. The circular polarization component tends to continuously rotate the rod with a torque of $M_\text{C}$.  (d) The nanorod thus experiences a rotational tilted washboard potential energy landscape as a function of orientation angle $\varphi$. The potential barrier height in the preferred rotation direction can be calculated to $\Delta U=0.8k_\text{B}T$ for parameters mimicking experimental data (nanorod with dimensions 65 nm$\times$147 nm, $\lambda_\text{Laser}$=830 nm, $P_\text{Laser}$=6 mW, $T$=326 K) for the case when the elliptically polarized wave has $\Delta\phi=40^\circ$. The barrier height in the opposite rotation direction is 9.4 $k_\text{B}T$.}
\end{figure}

\subsection{\label{sec:level2}Transition from continuous rotation to stochastic jumps of an individual gold nanorod}

\begin{figure*}
\includegraphics{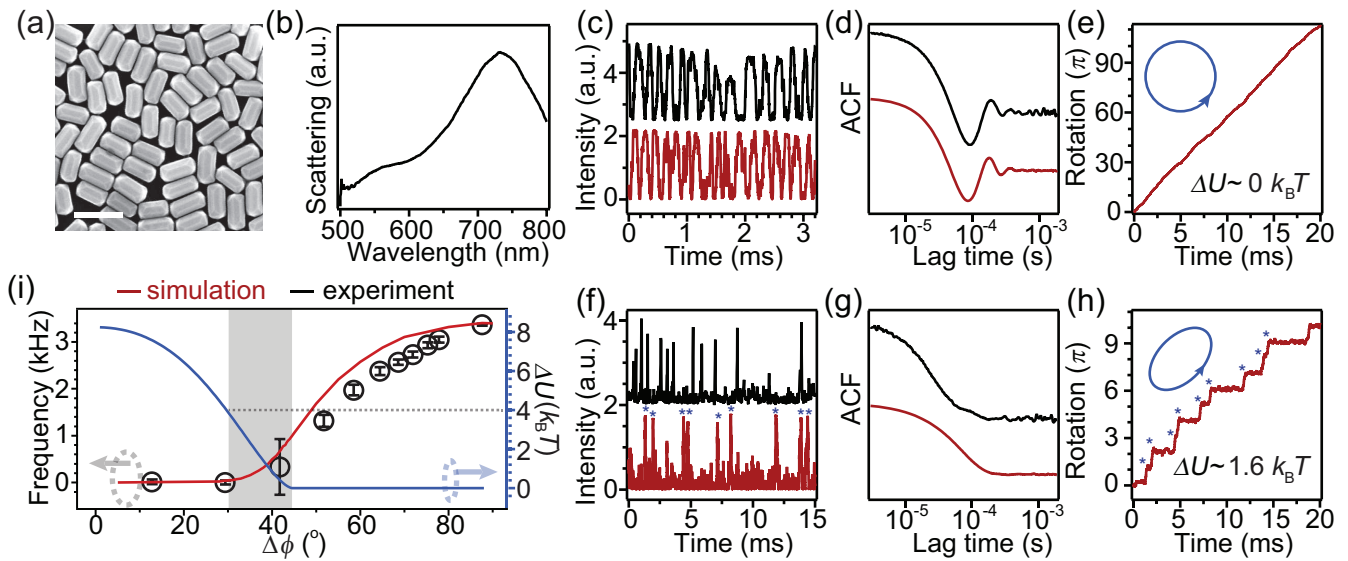}
\caption{\label{figure2} Transition from continuous rotation to discrete jumps of gold nanorods. (a) Scanning electron microscopy image of the nanorods (scale bar 200 nm). (b) Scattering spectrum of a trapped nanorod. The strong peak at $\sim$740 nm is caused by the long-axis surface plasmon resonance of the nanorod. (c$-$h) Measured (black) and simulated (red) rotational dynamics of the rod undergoing continuous rotation (c$-$e) in the presence of an almost perfectly circularly polarized laser field ($\Delta\phi=88^\circ$) and discrete rotational jumps (f$-$h) due to an elliptically polarized field ($\Delta\phi=37^\circ$). Figure (e,h) shows the nanorods orientation angle $\varphi$ versus time while (c,d) and (f,g) shows the corresponding scattering intensity time traces and intensity autocorrelation functions (ACFs), respectively. The data is based on measurements and simulations of cross-polarized backscattering from the nanorod. The blue stars in (f) and (h) mark out individual jumps. (i) Polarization-dependent rotation of a nanorod. We showed both measured (black) and calculated (red) rotational frequency of a nanorod as the laser polarization continuously varies from almost circular to linear. The blue curve indicates the barrier height at varying $\Delta\phi$. In the highlighted area ($30^\circ<\Delta\phi<45^\circ$), where the potential barrier $\Delta U\in(0, 4k_\text{B}T)$, we observe the nanorod undergoing a transition from continuous rotation to discrete jumps. The nanorod stops continuous rotation when $\Delta\phi$ decreases to around $40^\circ$ where an effective rotational potential barrier $\Delta U\approx k_\text{B}T$.}
\end{figure*}

We studied the rotational dynamics of gold nanorods optically trapped in 2D against a cover glass in an optical tweezers setup based on an 830 nm laser beam with tunable polarization ellipticity (see Methods in the Supplemental Material \footnotemark[\value{footnote}]). The nanorods had an average size of $(147\pm10\ \text{nm})\times(65\pm5\ \text{nm})$ (Fig.~\ref{figure2}a) and were prepared by a seed-mediated growth method \cite{chen2013gold,shao2015gold}. The exemplary dark-field scattering spectrum of an individual trapped nanorod (Fig.~\ref{figure2}b) shows one weak surface plasmon resonance at around 550 nm and a strong mode at around 740 nm overlapping the 830 nm trapping laser wavelength. The relative strengths of the resonance peaks correspond to a polarizability along the long-axis that is more than an order-of-magnitude higher than along the short axis at a wavelength of 830 nm, thus confirming the assumption of an essentially 1D polarizability tensor. The attractive and plasmon-enhanced optical gradient force keeps the particle trapped and aligned in the laser focus $xy$-plane while the Coulomb repulsion from the cover glass and the radiation pressure prevent the particle from escaping the trap along the $z$-axis \cite{shao2015gold}.

We first tracked the rotational dynamics of a trapped nanorod by analyzing the back-scattered laser light $I_\text{sca}^\text{P}$ from the particle using polarization selective detection in which a polarizer oriented perpendicular to the linear polarization component $E_\text{L}$ of the trapping beam has been placed in front of the detector. The nanorod angle variations are thus converted to fluctuations in $I_\text{sca}^\text{P}$ because of the nanorods' highly polarized scattering properties \cite{shao2015gold}. $I_\text{sca}^\text{P}(t)$ shows a periodic oscillation with superimposed fluctuations due to rotational Brownian motion for the case of an almost circularly polarized trapping field ($\Delta\phi=88^\circ$, Fig.~\ref{figure2}c, black trace). The corresponding autocorrelation function $C(\tau)$ of $I_\text{sca}^\text{P}(t)$ (Fig.~\ref{figure2}d, black trace) can be analyzed using $C(\tau)=I_0^2+0.5I_1^2\exp(-\tau/\tau_0)\cos(4\pi f\tau)$ \cite{shao2015gold}, which yields the nanorod average rotation frequency as $f=2460\pm20$ Hz and the autocorrelation decay time as $\tau_0=103\pm2$ $\mu\text{s}$. The nanorod ceased to rotate when we switched the laser polarization to elliptical ($\Delta\phi=37^\circ$), as is evident from the lack of a well-defined periodicity in the measured $I_\text{sca}^\text{P}(t)$ and $C(\tau)$ (Fig.~\ref{figure2}f and g, black traces). However, the recorded intensity trace nevertheless exhibits distinct occasional burst. We interpret these features as due to well-defined but stochastic thermal jumps in nanorod orientation.

Next, stochastic simulations were performed to gain further insight into the rotation process. The Brownian dynamics of a nanorod trapped in the tilted washboard potential $U(\varphi)$ can be simulated using the equation-of-motion \cite{Risken1996,shao2015gold}:
\begin{equation}
J\ddot{\varphi}=-\gamma_\text{r}\dot{\varphi}-\cos\varphi\sin\varphi\cdot\text{Re}(\alpha)E_\text{L}^2+\frac{1}{2}\text{Re}(\alpha)E_\text{C}^2+\xi(t).
\label{eq:two}
\end{equation}
Here, $J$ is the nanorod moment of inertia and the first three terms on the right-hand side represent, respectively, a viscous damping torque, characterized by a rotational friction coefficient $\gamma_\text{r}$, the restoring torque due to the linear polarization component $E_\text{L}$ and the driving torque due to the circular polarization component $E_\text{C}$. The last term represents a stationary Gaussian noise torque with zero mean and autocorrelation function $\langle\xi(t)\xi(0)\rangle=2\gamma_\text{r}k_\text{B}T_\text{r}\delta(t)$, where $T_\text{r}$ is the effective temperature for rotational Brownian motion \cite{hajizadeh2017brownian}. The temporal variation in nanorod orientation $\varphi(t)$ obtained from Eq.~(\ref{eq:two}) can in turn be used to calculate $I_\text{sca}^\text{P}(t)$ and $C(\tau)$ for comparison with experiments.

Fig.~\ref{figure2}c$-$h (red traces) shows simulation results for almost circular ($\Delta\phi=88^\circ$) and elliptical ($\Delta\phi=37^\circ$) polarization using simulation parameters selected to match experimental conditions, including a fixed $T_\text{r}=320$ K estimated \cite{shao2015gold,hajizadeh2017brownian} from the experimental $\tau_0$ (see the Supplementary Material for details \footnotemark[\value{footnote}]). For the circular polarization case, the simulated $\varphi(t)$ evolves continuously (Fig.~\ref{figure2}e), corresponding to continuous rotation, and the calculated $C(\tau)$ yields $f=2643\pm7$ Hz and $\tau_0=100\pm1$ $\mu$s in excellent agreement with the experimental results. For the elliptical polarization case, $\varphi(t)$ instead exhibits a staircase behavior (Fig.~\ref{figure2}h) corresponding to discrete and random $\pi$ jumps in one direction given by the driving torque, separated by periods of almost fixed alignment along the linear polarization direction. The resulting intensity trace and autocorrelation function are again in good agreement with the experimental observations Fig.~\ref{figure2}f$-$g). Thus, the simulations confirm that thermal agitation occasionally forces the nanorod to jump out of the local minima of the washboard potential to the next, lower, potential well, resulting in discrete intensity bursts in $I_\text{sca}^\text{P}$ that can be tracked experimentally.

We further varied the polarization state of the trapping laser continuously. Both experiments and simulations showed that the gold nanorod rotation becomes increasingly slow as the polarization becomes increasingly elliptical. When $\Delta\phi$ decreases to $\sim40^\circ$, the oscillating feature in the intensity autocorrelation function disappears (Fig.~\ref{figure2}i; more details can be found in the Supplementary Material \footnotemark[\value{footnote}]), suggesting that a barrier exists and stops the nanorod's continuous rotation. However, the nanorod undergoes stochastic rotational jumps from time to time and we can still calculate an effective rotation frequency by counting the number of discrete jumps. When the polarization becomes even more elliptical, the barrier in the rotation potential becomes high enough to keep the rod aligned with the major axis of the polarization ellipse. The nanorod thus exhibits a `rotation' frequency asymptotic towards zero. From simulation, we further observed that the effective rotational diffusion of the nanorod varies as the laser polarization changes, similar to the reported result of translational diffusion of a Brownian particle on tilted washboard potentials \cite{reimann2001giant}. Interested readers can find more discussion in the Supplementary Material \footnotemark[\value{footnote}].

\subsection{\label{sec:level2}Stochastic jump dynamics of gold nanorods trapped by elliptical polarization}

\begin{figure}
\includegraphics{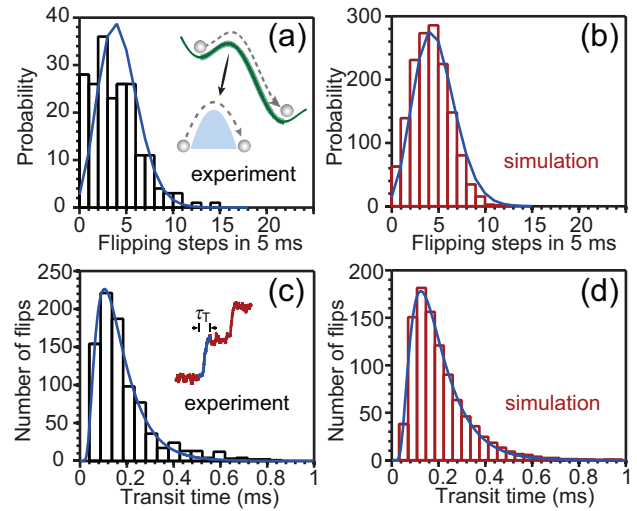}
\caption{\label{figure3} Stochastic rotational jump dynamics of a gold nanorod. The gold nanorod was trapped using an elliptically polarized laser beam with $\Delta\phi=37^\circ$ and a power of 6 mW. (a$-$d) Experimentally measured and simulation-calculated probability distributions of nanorod flips in a fixed time interval of 5 ms (a,b) and transit times of such jump (c,d). The nanorod rotational effective Brownian temperature $T_\text{r}$ was set at 360 K in simulation to achieve good agreement with experiments. The laser trap forms a tilted washboard rotation potential with a barrier $\Delta U=1.4\ k_\text{B}T_\text{r}$, and near a local maximum the potential landscape can be modeled with an inverted harmonic potential (inset in a). The rod was agitated thermally to overcome this barrier to jump by an angle of $\pi$, and each jump process was recorded by one scattering intensity peak shown in Fig.~\ref{figure2}f. The numbers of jumps follow Poisson distribution (blue curves in a and b). The full distributions of transit times are well fitted by formula derived from Kramers theory (blue curves in c and d), with the coefficients of determination $R^2=$ 0.979 and 0.994, respectively.}
\end{figure}

To test quantitatively the physical attributes of nanorod stochastic jumps, including their rate and transit time, we further examined the measured scattering signals and the simulation results to discern further details. Statistics on both the experiment and simulation results revealed that the number of nanorod flips in a certain time interval $X$ follows a Poisson probability distribution $P(X=S)=e^{-\lambda} \lambda^S/S!$ (Fig.~\ref{figure3}a,b), with the Poisson mean $\lambda$ determined by the potential barrier height relative to $k_\text{B}T_\text{r}$. When the time interval is set at 5 ms, the fitting-obtained $\lambda$ are $4.3\pm0.3$ for experiment and $4.7\pm0.1$ for simulation (laser power 6 mW, $\Delta\phi=37^\circ$), respectively.

Additional information can be obtained through studying the duration of each individual stochastic jump, as the average value and the variability in transit times reflect kinetics and the fundamentally statistical nature of the stochastic jump process. The transit time $\tau_\text{T}$ is much shorter than the first-passage time defined as how long it takes for the nanorod to rotate by $\pi$. This is very different from the case when the nanorod undergoes continuous rotation (more details can be found in the Supplementary Material \footnotemark[\value{footnote}]). $\tau_\text{T}$ was found to vary widely, from less than 80 $\mu$s to over 600 $\mu$s (Fig.~\ref{figure3}c,d), with average values $\langle\tau_\text{T}\rangle=158\pm114$ $\mu$s (experiment) and $117\pm47$ $\mu$s (simulation). Additionally, both experimental and simulation results revealed that the broadly distributed $\tau_\text{T}$ has a peak at around 120 $\mu$s and a long exponential tail (Fig.~\ref{figure3}c,d). This behavior is similar to that expected for transit across harmonic barriers in the high-barrier limit ($\Delta U>>k_\text{B}T$) in the Kramers regime \cite{kramers1940brownian,RevModPhys.62.251,chaudhury2010harmonic}. Specifically, in our case, the potential landscape near a local maximum can be modeled with an inverted harmonic potential with a ``spring constant'' $\kappa_\text{b}$, $V(\varphi)\approx V_\text{max}-\kappa_\text{b}(\varphi-\varphi_\text{max})^2/2$ (schematic in Fig.~\ref{figure3}a inset). When the transition region is from $(\varphi_\text{max}-\varphi_0)$ to $(\varphi_\text{max}+\varphi_0)$, the barrier height $\Delta U=\kappa_\text{b}\varphi_0^2/2$ and the transit time $\tau_\text{T}$ has a distribution $P(\tau_\text{T})$. For the one-dimensional diffusion model determined by  $J\ddot{\varphi}=-\gamma_\text{r}\dot{\varphi}-V'(\varphi)+\xi(t)$, $P(\tau_\text{T})$ is predicted to have the form \cite{chaudhury2010harmonic,doi:10.1063/1.2434966}:
\begin{equation}
P(\tau_\text{T})=\frac{\omega_\text{K}\sqrt{\Delta U/(k_\text{B}T)}}{1-\text{erf}[\sqrt{\Delta U/(k_\text{B}T)}]}\frac{\exp[-\Delta U\coth(\omega_\text{K}\tau_\text{T}/2)/(k_\text{B}T)]}{\sinh(\omega_\text{K}\tau_\text{T}/2)\sqrt{2\pi\sinh(\omega_\text{K}\tau_\text{T})}}
\label{eq:three}
\end{equation}
The distribution in Eq.~(\ref{eq:three}) decays exponentially for large $\tau_\text{T}$ as $P(\tau_\text{T})\approx2\omega_\text{K}[\Delta U/(k_\text{B}T)]\text{exp}(-\omega_\text{K}\tau_\text{T})$. The parameter $\omega_\text{K}$ sets the time scale for decay away from states near the top of the barrier, $\omega_\text{K}=\kappa_\text{b}/\gamma_\text{r}$.

In both experiment and simulation, $P(\tau_\text{T})$ are well fitted by Eq.~(\ref{eq:three}) (Fig.~\ref{figure3}c,d). The barrier heights  $\Delta U$ returned by the fit are $1.10\pm0.30\ k_\text{B}T$ (experiment) and $1.36\pm0.08\ k_\text{B}T$ (simulation), both in good agreement with the value calculated from the laser polarization according to Eq.~(\ref{eq:one}): $\Delta U=1.4\ k_\text{B}T_\text{r}$. The values of fitting-obtained $\omega_\text{K}$ are $1.13\pm0.17\times10^4\ \text{s}^{-1}$ (experiment) and $1.02\pm0.03\times10^4\ \text{s}^{-1}$ (simulation). Given that the rotational diffusion constant $D_\text{r}$ is determined by $D_\text{r}=k_\text{B}T/\gamma_\text{r}$, we can write $\omega_\text{K}$ as well as the average transit time in terms of this quantity through
\begin{subequations}
\label{eq:whole1}
\begin{equation}
 \omega_\text{K}=\kappa_\text{b}/\gamma_\text{r}=D_\text{r}\kappa_\text{b}/(k_\text{B}T),
\label{subeq:1}
\end{equation}
\begin{equation}
\langle\tau_\text{T}\rangle=\ln\big(2e^\gamma\Delta U/(k_\text{B}T)\big)/\omega_\text{K},
\label{subeq:2}
\end{equation}
\end{subequations}
where $\gamma$ is Euler's constant \cite{chaudhury2010harmonic}. Given that $\kappa_\text{b}$ can be determined by fitting the energy landscape to be $\kappa_\text{b}=3.4\pm0.1\ k_\text{B}T/\text{rad}^2$ at $\Delta\phi=37^\circ$, we can calculate the rotational diffusion coefficient $D_\text{r}$ from $\omega_\text{K}$ and $\langle\tau_\text{T}\rangle$ according to Eqs.~(\ref{eq:whole1}). For the simulation result, $D_\text{r}$ calculated from $\omega_\text{K}$ is $3.0\pm0.1\times10^3\ \text{s}^{-1}$, close to $D_\text{r}=4.0\pm1.6\times10^3\ \text{s}^{-1}$ calculated from $\langle\tau_\text{T}\rangle$. In experiment, $D_\text{r}$ is calculated from $\omega_\text{K}$ and $\langle\tau_\text{T}\rangle$ to be $3.3\pm0.5\times10^3\ \text{s}^{-1}$ and $3.0\pm2.2\times10^3\ \text{s}^{-1}$, respectively, which are also in good agreement with each other. The values of $D_\text{r}$ calculated from measured $\omega_\text{K}$ and $\langle\tau_\text{T}\rangle$ are close to the result directly calculated by modeling the nanorod as a prolate ellipsoid in water ($D_\text{r}=4.6\times10^3\ \text{s}^{-1}$), validating Kramers description of the nanorod rotational jump transition.

\begin{figure}
\includegraphics{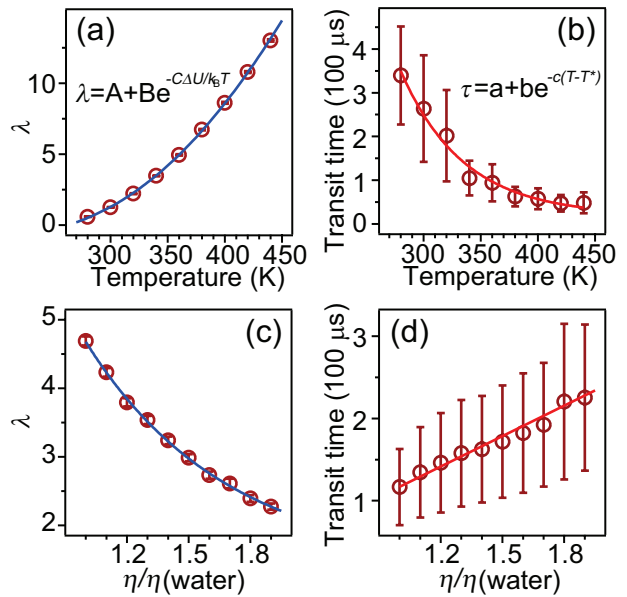}
\caption{\label{figure4} Temperature- and medium viscosity-dependent nanorod stochastic jump dynamics. (a,b) Rotational effective Brownian temperature $T_\text{r}$-dependent average number of flips in a time interval of 5 ms (a) and transit time (b) calculated from simulation (red dots). The blue curve in (a) and the red curve in (b) are fitting results, with the coefficients of determination $R^2=$ 0.975 and 0.998. (c,d) Medium viscosity $\eta$-dependent average number of flips (c) and transit time (d) calculated from simulation (red dots). The blue curve in (c) and the red curve in (d) are exponential and linear fitting results, with the coefficients of determination $R^2=$ 0.999 and 0.968.}
\end{figure}

Furthermore, the rates and transit times of the nanorod stochastic jumps are highly dependent on the temperature and viscosity of the local nanoenvironment. If we artificially increase $T_\text{r}$ in simulation, the calculated Poisson distribution means $\lambda$ and the average transit time $\langle\tau_\text{T}\rangle$ exhibit rapid increase and decrease (Fig.~\ref{figure4}a,b), respectively. $\lambda$ indicating the transition rate follows an exponential (Boltzmann) dependence on the energy barrier height $\Delta U$ (Fig.~\ref{figure4}a). $\langle\tau_\text{T}\rangle$ decreases exponentially with $T_\text{r}$ (Fig.~\ref{figure4}b). $\lambda$ and $\langle\tau_\text{T}\rangle$ are also very sensitive to the medium viscosity $\eta$. As $\eta$ increases, $\lambda$ decreases exponentially and $\langle\tau_\text{T}\rangle$ increases linearly (Fig.~\ref{figure4}c,d), according to our simulation results. The sensitive temperature and viscosity dependence of the nanorod stochastic jump dynamics implies that the gold nanorod manipulated by an elliptical polarization can work as a sensing element to probe the local temperature and viscosity in solution.

\section{Discussion}
We have shown the construction of tilted washboard rotational potentials for optically trapped Brownian plasmonic gold nanorod motors, by rather simple means, utilizing elliptical polarizations. The gold nanorods were modulated from continuous rotation to discrete jumps by simply adjusting the polarization state of the trapping laser. In addition, we studied individual stochastic jump processes of the nanorod at critical laser polarization, finding that the jump dynamics is in good agreement with that predicted by Kramers theory. Our measurement results can be directly used to help understand mechanisms in molecular motors \cite{michl2009molecular} and rotating dipoles in external fields \cite{praestgaard1981model}.

The plasmonic nanorod trapped by elliptical polarization is a simple optical analogy to many other physical systems. It provides a powerful tool for investigating fundamental questions where the problem of Brownian motion in tilted periodic potentials arises \cite{rice2013advances,Risken1996,fulde1975problem,stratonovich1967topics}. The colloidal nanoparticle trapped in solution works in an overdamped regime; one can extend this to an underdamped regime by trapping plasmonic nanoparticles in air or in vacuum \cite{rondin2017direct,jauffred2015optical}. The plasmonic nanorod optically trapped with elliptical polarization thus generates a universal model system for the nonequilibrium thermodynamics problem of “Brownian diffusion over periodic barriers”. The small size ($\sim$100 nm) and short characteristic time scale ($\sim100\ \mu$s) of the Brownian nanorod allow for fast statistical investigations. Experimental parameters can be controlled and varied \textit{in situ} easily in an optical way, significantly reducing the complexity, incompatibility, and inadaptability of other physical systems conventionally employed. As we have shown, a very simple and well known theoretical model can be utilized to capture the system dynamics \cite{Risken1996,coffey2004langevin}. This, in turn, means that it is also straight-forward and robust to extract parameters by model fitting. As a result, we believe that our study facilitates the use of Brownian plasmonic nanoparticles as new probes to study fundamental issues with broad interest, such as giant acceleration of particle diffusion \cite{reimann2001giant}, connection between statistical physics and information theory \cite{berut2012experimental,parrondo2015thermodynamics}, and hydrodynamic synchronization \cite{koumakis2013stochastic}.

In addition, we have realized the full rotational control of the light-driven gold nanorod motors. The elliptical polarization trapped gold nanorod can also work as a ratchet that harvests overdamped Brownian noise and rectifies the Brownian motion at thermal non-equilibrium \cite{wu2016near}. Moreover, by combining the structure Brownian dynamics analysis, which can probe the local viscosity and temperature, and the plasmonic molecular analysis techniques such as refractometric sensing \cite{zijlstra2012optical} and surface-enhanced Raman scattering \cite{xu1999spectroscopy}, the optical-potential-controlled gold nanorod further becomes a multifunctional sensing platform to probe different characteristics of local nanoenvironment \cite{andren2017probing,vsipova2018photothermal}.

\begin{acknowledgments}
The authors thank Nils Odebo L{\"a}nk for help with the FDTD simulations. Support and advice by Prof. Giovanni Volpe and Prof. Andreas Isacsson is gratefully acknowledged. This work was supported by the Knut and Alice Wallenberg Foundation.
\end{acknowledgments}

\nocite{*}

\bibliography{shao2018prb}% Produces the bibliography via BibTeX.

\end{document}